%% file: main.tex
\documentclass[conference]{IEEEtran}
\IEEEoverridecommandlockouts
\usepackage{cite}
\usepackage{amsmath,amssymb,amsfonts}
\usepackage[detect-all,group-separator={,}]{siunitx}
\usepackage{algorithmic}
\usepackage{graphicx}
\usepackage{textcomp}
\usepackage{xcolor}
\usepackage{multirow}
\def\BibTeX{{\rm B\kern-.05em{\sc i\kern-.025em b}\kern-.08em
    T\kern-.1667em\lower.7ex\hbox{E}\kern-.125emX}}

\input{misc}

\begin{document}

\title{FastqZip: An Improved Reference-Based Genome Sequence Lossy Compression Framework}

\author{\IEEEauthorblockN{
Yuanjian Liu\IEEEauthorrefmark{2}\IEEEauthorrefmark{1},
Huihao Luo\IEEEauthorrefmark{2},
Zhijun Han\IEEEauthorrefmark{2},
Yao Hu\IEEEauthorrefmark{2},
Yehui Yang\IEEEauthorrefmark{2},\\
Kyle Chard\IEEEauthorrefmark{1},
Sheng Di\IEEEauthorrefmark{3},
Ian Foster\IEEEauthorrefmark{1}\IEEEauthorrefmark{3},
Jiesheng Wu\IEEEauthorrefmark{2}
}
\IEEEauthorblockA{\IEEEauthorrefmark{2}
Alibaba Cloud, China}
\IEEEauthorblockA{\IEEEauthorrefmark{1}
University of Chicago, Chicago, IL, USA}
\IEEEauthorblockA{\IEEEauthorrefmark{3}
Argonne National Laboratory, Lemont, IL, USA
}

yuanjian@uchicago.edu, jiesheng.wu@alibaba-inc.com
}

\maketitle

\thispagestyle{plain}
\pagestyle{plain}

\begin{abstract}
Storing and archiving data produced by next-generation sequencing (NGS) is a huge burden for research institutions. Reference-based compression algorithms are effective in dealing with these data. Our work focuses on compressing FASTQ format files with an improved reference-based compression algorithm to achieve a higher compression ratio than other state-of-the-art algorithms. We propose FastqZip which uses a new method mapping the sequence to reference for compression, allows reads-reordering and lossy quality scores, and the BSC or ZPAQ algorithm to perform final lossless compression for a higher compression ratio and relatively fast speed. Our method ensures the sequence can be losslessly reconstructed while allowing lossless or lossy compression for the quality scores. We reordered the reads to get a higher compression ratio. We evaluate our algorithms on five datasets and show that FastqZip can outperform the SOTA algorithm Genozip by around 10\% in terms of compression ratio while having an acceptable slowdown.
\end{abstract}


\input{sections/1-introduction}

\input{sections/2-related-work}

\input{sections/3-problem-formulation}

\input{sections/4-methods}

\input{sections/5-evaluations}

\input{sections/6-conculsions}

\section*{acknowledgment}


We would like to extend our gratitude to BGI Genomics for sharing their public datasets for our benchmark. We are also grateful to the University of Chicago for providing academic advice. The collaboration among Alibaba Cloud, BGI Genomics, and the University of Chicago was fundamental to the successful execution of this study, and we are sincerely thankful to all parties involved.



\bibliographystyle{IEEEtran}

\IEEEtriggeratref{24}

\bibliography{
    refs/genome,
    refs/compression
}

\end{document}

%% file: misc.tex
\usepackage{todonotes}

%% file: sections/1-introduction.tex
\section{Introduction}
\label{sec:intro}

Next-generation sequencing (NGS) technologies continue to improve their performance. For example, the DNBSQ-T20 sequencing platform released by Complete Genomics at the 2023 Conference on Genome Biology and Technology can generate 22 TB of sequence data per day. The high cost of storing these data means that even modest levels of compression can be of great benefit. 

Raw sequencing data are typically stored in FASTQ format~\cite{fastq-format}. A FASTQ file consists of a separate entry for each short sequence, consisting of four lines: an identifier string, a nucleotide sequence (the read), the character '+', and quality scores. The identifier string contains information about the sequencing technology and other metadata obtained from the sequencing machine, which uniquely describes a read. The nucleotide sequence is a string of A, C, G, T, and N characters representing the bases (base-pairs) of the DNA sequence. There are some other characters for storing protein sequences, which do not exist in our datasets. The quality scores record the confidence of each base generated by the sequencing machine. Due to their higher entropy and larger alphabets, quality scores have proven to be more difficult to compress than reads~\cite{hernaez2019genomic}.

To reduce the space required to store FASTQ files, researchers focus on the compression of the reads and quality scores, which consume most of the space and carry the most relevant information~\cite{hernaez2019genomic}.
Traditional general-purpose compression algorithms such as gzip~\cite{gzip}, and bzip2~\cite{bzip2} fail to obtain a high compression ratio for sequencing data.
Thus, many specialized FASTQ compressors\cite{reference-compression,path-encoding-ref,RENANO,genozip} have been proposed. The most successful of these algorithms are so-called reference methods, which exploit the fact that more than 99\% of human genomes are identical~\cite{initial-genome} to reduce greatly the storage space required. 

\begin{figure}[thb]
    \centering
    \includegraphics[width=1\linewidth]{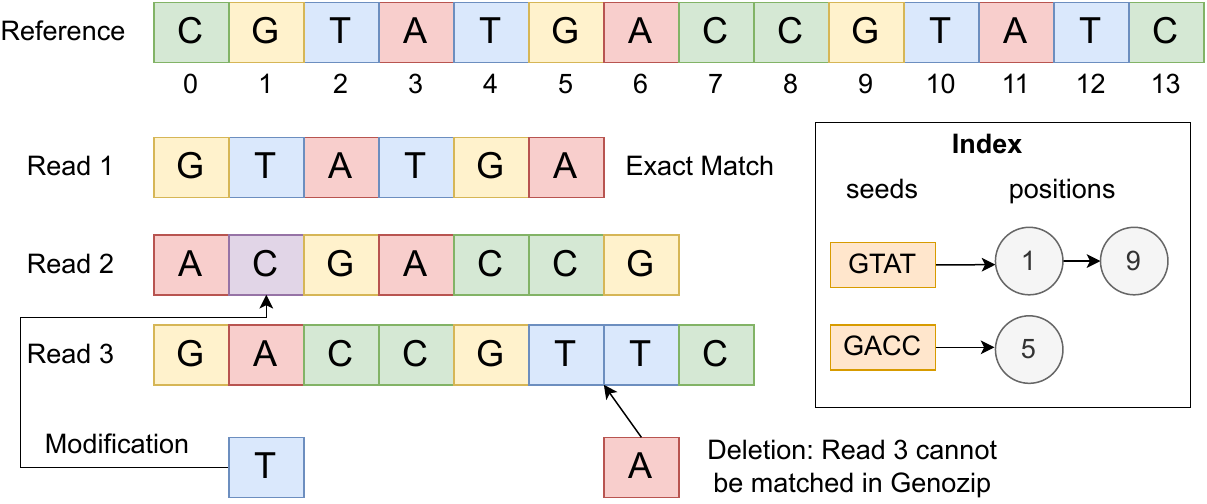}
    \caption{Reference-based sequence matching process: (1) use seeds to build an index for the reference sequence; (2) find matching locations for reads on the reference sequence; (3) for unmatched bases, store the difference. Our algorithm performs better matching by storing more seeds for a higher chance of matching, and local search for insertion and deletion detection.}
    \label{fig:ACGT-matching}
\end{figure}

Most existing genome sequence compression algorithms\cite{genozip,seq-compress,genome-compress} are lossless, a choice that limits achievable compression but is not necessary for most downstream analyses. In practice, 1) the reads in a FASTQ file are usually independent of each other, and thus can be reordered with no impact on most downstream analysis\cite{ngs-application}; 2) the informtion contained in a read's (verbose) identifier is typically ignored, and thus can be replaced with a more compact unique identifier; and 3)  analysis scores are often not sensitive to quality scores, and thus scores can be quantized to improve compression performance. 

In this work, we propose an algorithm that employs more fine-grained read-to-reference matching, read sorting, and optional lossy quality compression to achieve a compression ratio that is higher than the state-of-the-art (SOTA) algorithms for raw sequencing data. 
Our key contributions are as follows:
\begin{enumerate}
    \item We propose a reference-based genome sequence compression framework, FastqZip, which outperforms SOTA compression algorithms and achieves a higher compression ratio.
    \item We design a novel sequence matching procedure that can find a match when the Hamming Distance is huge but the Edit Distance is small between the read and the reference. Therefore, many previously unmatchable reads can be reconstructed from the reference sequence.
    \item We conduct comprehensive evaluations and compare the performance of our algorithms with SOTA algorithms. The result shows that FastqZip has the best compression ratio over five common datasets.
    \item We show that our algorithm scales better than existing algorithms when parallelized over many resources. 
\end{enumerate}

The rest of the paper is organized as follows. In Section~\ref{sec:related-work}, we review previous work on genome sequence compression. In Section~\ref{sec:problem-form}, we formally define the problem. In Section~\ref{sec:methods}, we introduce FastqZip and our algorithms. In Section~\ref{sec:evaluation}, we conduct evaluations on five real-world datasets and compare our algorithm with SOTA methods. In Section~\ref{sec:conclusion}, we conclude and discuss potential future directions.

%% file: sections/2-related-work.tex
\section{Related Work}
\label{sec:related-work}

\subsection{DNA Sequencing Technologies}

The Human Genome Project\cite{human-genome-project} sought to sequence the DNA of every human chromosome. Early efforts to sequence genes such as \cite{maxam-gilbert} are painstaking, time-consuming, and labor-intensive. Sanger sequencing\cite{sanger-sequencing} changed this situation by using a purified DNA polymerase enzyme to synthesize DNA chains of varying lengths. The procedure is sensitive enough to distinguish DNA fragments with just a single nucleotide difference. Continued developments have produced next-generation sequencing (NGS)  methods and machines that generate data at rates that have increased faster than Moore's Law. 

Illumina\cite{illumina}, a widely used modern DNA sequencing platform, produces raw data files in binary format (BCL) that contain the fluorescent signal intensities for each nucleotide incorporation event. To convert this data into a more useful format, the instrument software performs base calling and quality scoring to generate a FASTQ\cite{fastq-format} file. 

\subsection{DNA Sequence Compression Algorithms}

Existing FASTQ sequence compression algorithms can be categorized into two classes: reference-based and reference-free algorithms. \textit{Reference-based algorithms} map the nucleotide sequences in a FASTQ file to a reference genome and use the mapped positions to encode the sequences. Examples include LW-FQZip\cite{LW-FQZip}, LW-FQZip2\cite{LW-FQZip2}, GTZ\cite{gtz}, and genozip\cite{genozip}. 

\textit{Reference-free algorithms} are used when a reference genome is not available. For example, Leon\cite{LEON} and Quip\cite{Quip} use assembly-based algorithms, in which a De Bruijn graph is constructed from the already compressed sequences and incoming sequences are matched to the graph to locate the longest exact matches. Other specialized FASTQ compressors initially perform a form of transformation (read-identifier tokenization or 2-bit nucleotide encoding) followed by statistical modeling and entropy coding. Examples of such approaches are DSRC2\cite{DSRC2}, FQC\cite{FQC}, Fqzcomp\cite{fqzcomp}, Slimfastq\cite{fqzcomp}, LFQC\cite{LFQC}, and Spring\cite{Spring}. FQSqueezer\cite{FQSqueezer}, a more recent compressor, uses partial matching and a dynamic Markov coder.

Generally, reference-based compressors perform better in terms of both compression time and ratio than reference-free compressors, and thus we focus here on a reference-based algorithm for FASTQ compression. Moreover, we tested all of the aforementioned algorithms in Section~\ref{sec:evaluation} and found that most suffer from low compression ratio, extremely slow compression speed, and bad scalability on multiple processors, and furthermore are painstaking to build and can be erroneous on certain datasets. Therefore, we hope to provide a FASTQ compression tool that has better performance and is easier to use.

%% file: sections/3-problem-formulation.tex
\section{Problem Formulation}
\label{sec:problem-form}


We provide a formal definition of our reference-based genome compression problem.

A FASTQ file contains information about a set of reads, $R_i$, each defined by three components: a target sequence, a set of quality scores, and an identifier. In a raw FASTQ file, the target sequence and quality scores each take equally around 49\% of the storage space, while the identifiers take the rest \~2\%. Each component can be compressed independently. The target sequence has to be lossless, but the order can be relaxed. We use the reference-based matching algorithm for the target sequence while using some traditional lossy/lossless algorithms for the quality scores and the identifiers.

For a single read $R_i$ with target genome sequence $X^{N}$, then given $Y{^M}$ as the reference information, we define an encoder $f(\cdot, \cdot)$ by mapping $X^{N}$ to a byte sequence $B^{K}$ with relationship $B^{K} = f(X^{N}, Y^{M})$, where $N$ is the target sequence length, $M$ is the reference sequence length, and $K$ is the compressed byte sequence length. One measure of a successful compressor is that it yields a $B^{K}$ for which $K<N$. The decoder $g(\cdot, \cdot)$ should then recover $B^{K}$ to $X^{N}$ with a function $g(B^{K}, Y^{M})$. Thus the encoder-decoder pair together recreate the original sequence:

\begin{equation}
    X^{N} = g(f(X^{N}, Y^{M}), Y^{M})
    \label{eq:encode-decode}
\end{equation}

We preserve the losslessness for each read but relax the order restriction for a group of reads in one FASTQ file. Given $K$ reads, each of length $N$, then after compression and decompression we have:
\begin{equation}
\begin{bmatrix}
    X^{N}_{i_1} \\
    X^{N}_{i_2} \\
    \vdots    \\
    X^{N}_{i_K} \\
\end{bmatrix} = G\left(F\left(\begin{bmatrix}
    X^{N}_{1} \\
    X^{N}_{2} \\
    \vdots    \\
    X^{N}_{K} \\
\end{bmatrix}, Y^{M}\right), Y^{M}\right)
\label{eq:order-relax}
\end{equation}
where $F$, $G$ are the corresponding functions of $f$, $g$ that can handle a vector of $X$, and each $X_{k}^N$ should be identical to one $X_{i_j}^N$, i.e., $\forall k \in [1, K], X_{k}^N, \exists i_j \in [1, K] $ such that $X_{k}^N = X_{i_j}^N$.

Moreover, there is a computation cost for $F(\cdot,\cdot)$ and $G(\cdot, \cdot)$. We use $T_f$ and $T_g$ to denote the time cost for the encoding and decoding process. Our algorithm should consider both the compression time cost and compression ratio, and therefore we define $s$(\emph{CR}, $T$\/) as a score function, where \emph{CR} is the compression ratio and $T$ is the (de)compression time.

The goal is to construct an encoding/decoding pair that maximizes $s(N/K, T_f + T_g)$ while preserving Equations~\ref{eq:encode-decode} and \ref{eq:order-relax}.

%% file: sections/4-methods.tex
\section{Methodology}
\label{sec:methods}

The FastqZip compression can be separated into four parts: (1) index loading, (2) sequence alignment, (3) sequence and quality segmentation, and (4) lossless compression. In this section, we will present the overall architecture and describe the core algorithm for each part, introduce the compressed file structure, and walk through the proposed parallel file structure for performance improvement.

\subsection{Compression Architecture}

\begin{figure}[htb]
    \centering
    \includegraphics[width=0.8\linewidth]{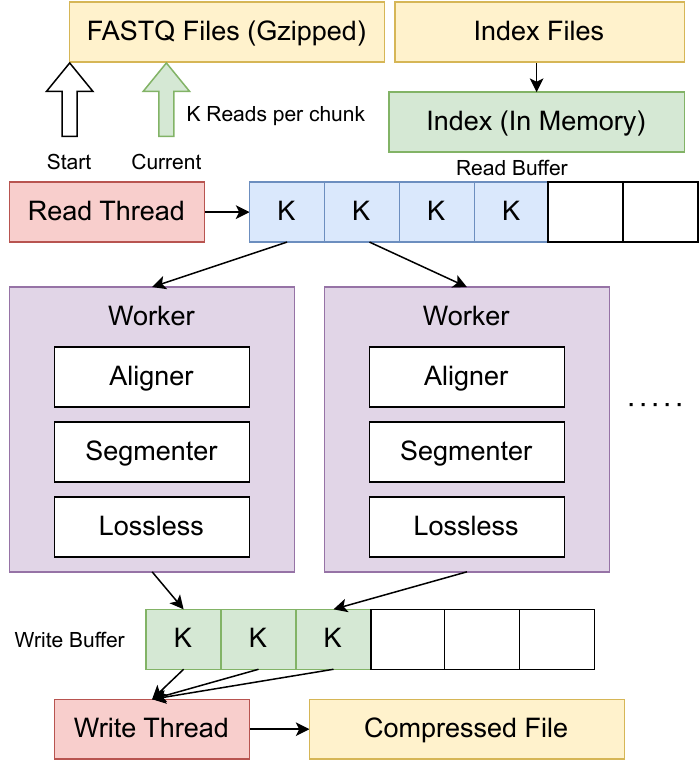}
    \caption{FastqZip compression architecture: The read thread must be sequential, but workers can proceed in parallel. The read buffer and write buffer allow maximum parallelism for the whole pipeline.}
    \label{fig:fastqzip-overall}
\end{figure}

We propose the compression architecture shown in \figurename~\ref{fig:fastqzip-overall}. The architecture employs one read thread, one write thread, and several worker threads to perform compression. These threads are synchronized by read and write buffers (in the program, the synchronization is performed by condition variables and mutexes). The read thread continues reading data from disks and stores them in the read buffer in memory. The worker will try to read a chunk from the read buffer. When the data is ready, one worker will retrieve it and remove the entry in the read buffer so that the read thread can read the next chunk. After the worker finishes compression, it puts the compressed data in the write buffer and tries to retrieve the next chunk from the read buffer. If the write buffer is full or the read buffer is empty, the worker thread waits. This design allows FastqZip to compress extremely large FASTQ files without breaking memory limits and to achieve high degrees of parallelism.


This architecture supports parallelism by allowing multiple workers to compress each chunk independently and write to the compressed file without waiting for any other workers. Moreover, we propose a compressed file structure that allows parallel reading which enhances the parallel decompression performance. However, because the read thread needs to read the gzipped file sequentially due to characteristics of the gzip algorithm, when there are more workers, the read thread can soon be too slow to fill up the read buffer. We will evaluate the parallelism and scalability later in Section~\ref{sec:evaluation}.

\subsection{Index Building \& Loading}

\begin{figure}[htb]
    \centering
    \includegraphics[width=0.9\linewidth]{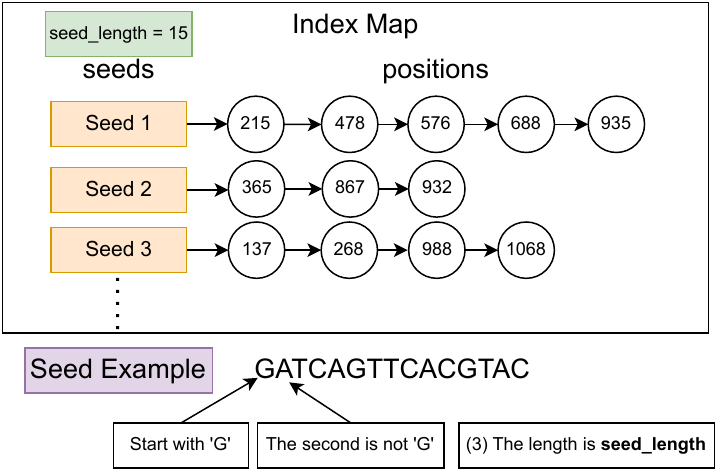}
    \caption{Index concept: we look for all valid seeds in the reference sequence and record their positions. There are multiple positions because the same seed may appear multiple times in different locations on the reference sequence.}
    \label{fig:index-map}
\end{figure}

An index is necessary for reference-based compression because naive long-string comparison is extremely slow. We only need to build the index once for each reference sequence. After the index is built, loading it into memory can be much faster during compression. Over the years, various data structures, such as hash tables and FM-index, have been used. We employ a key-value map in which short seed sequences serve as keys, and the values are the positions of those seeds in the reference sequence, as shown in \figurename~\ref{fig:index-map}. A seed is a short sequence that: (1) starts with the base `G', (2) has a second base that is not `G', and (3) has a predefined fixed length. The second condition is to avoid storing too many seeds of similar purpose when there are many continuous `G's on the reference sequence.

\begin{figure}[htb]
    \centering
    \includegraphics[width=1\linewidth]{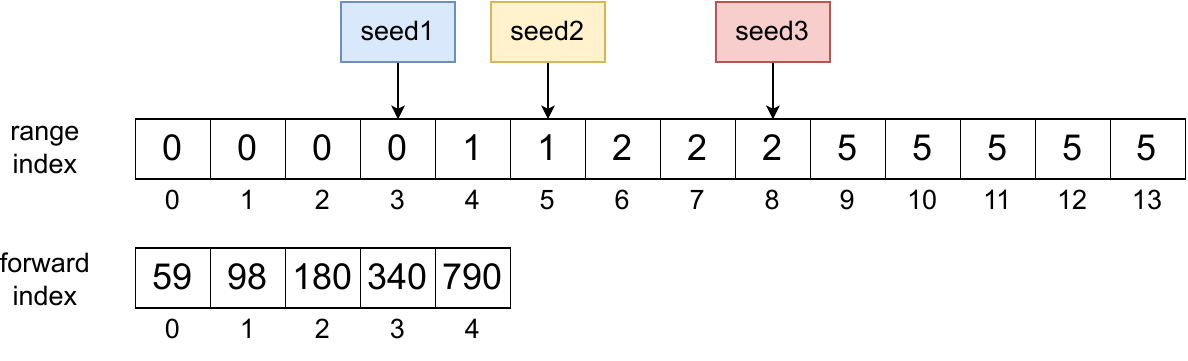}
    \caption{Index storage: The range index and forward index arrays together store the reference positions for all seeds. A seed can be uniquely mapped to an index $i$ in the index range array. The value in index\_range[$i$] is the starting index in the index forward array, and the value in index\_range[$i+1$] is the index after the ending index in the index forward array.}
    \label{fig:index-range}
\end{figure}

To simplify the storage of the index file, we propose three concepts: (1) forward sequence, (2) range index, and (3) forward index. The forward sequence is to connect the reference sequences to form one long sequence, and replace all non-ACGT bases with `A'. The range index is a fixed-length array storing the cumulative number of repetitions for seeds, as shown in \figurename~\ref{fig:index-range}. Each position in the range index is an integer converted from a seed. If we predefine the seed length to be 15, the index range's length will be $3 \times 4^{(15-2)}$, so any seed of length 15 can be uniquely mapped to one index in the index range array. The forward index is an array that stores the reference positions, following a strict order of each seed's converted integer. For example, in \figurename~\ref{fig:index-range}, seed1 appears once in the reference sequence, at position 59; seed2 also appears once, at position 98; and seed3 appears three times, at positions 180, 340, and 790.

Compared to prior work, Genozip\cite{genozip}, we store all the positions for repeated seeds instead of only four so that there can be a better match during the alignment step.

\subsection{Sequence Alignment}

Due to the high similarity among genome sequences of the same species, we can consider each read in a FASTQ file to be a short subsequence extracted from the reference sequence. The issue is that we do not know the exact position from which each read is extracted. Our goal in the sequence alignment step is to match each read to one position in the reference sequence so that, ideally, we can reconstruct the read with a position and the reference sequence. This task is complicated by mutations, insertions, deletions, and totally unmatchable reads. It is inefficient to test all possible locations, and other than exact matches, the algorithm needs to store some additional information and be smart to identify the match when there is some but not too much difference. In this section, we propose an improved global + local search approach that outperforms the existing algorithms\cite{genozip}\cite{seq-compress}\cite{genome-compress} in terms of compression ratio.

\begin{figure}[htb]
    \centering
    \includegraphics[width=1\linewidth]{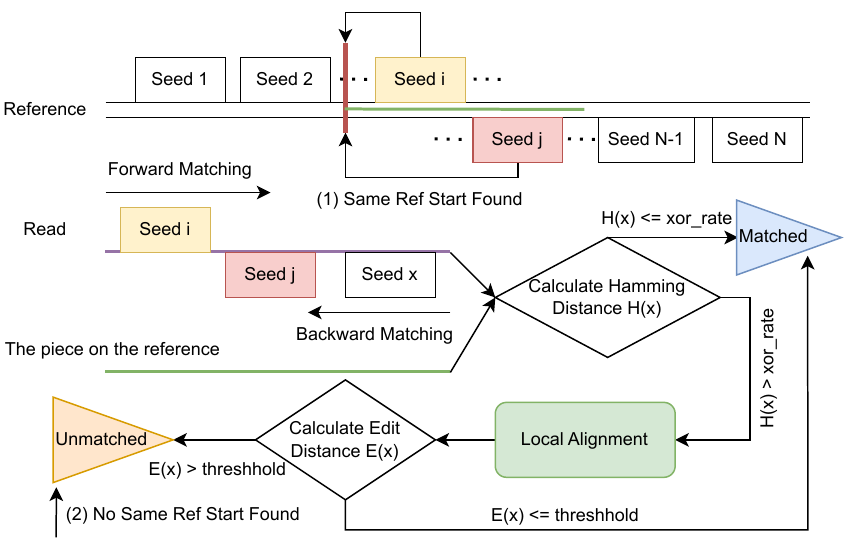}
    \caption{Alignment procedure: when multiple seeds exist on a single read, if a match exists, two seeds should match to the same starting position on the reference. If the candidate sequence on the reference has a very low Hamming Distance against the read, it is a match. If there are the same starting positions, but the Hamming Distance is large, we use our proposed local alignment to find a match with insertion or deletion.}
    \label{fig:alignment}
\end{figure}

The alignment procedure is illustrated by \figurename~\ref{fig:alignment}. For each read, we iterate through the seeds in both forward and backward directions and calculate the starting position of this read on the reference sequence. If two seeds appear to have the same starting position, it is likely that the read is indeed cut from the reference at that position. We consider this read a matchable candidate when the same reference start is found. We need to further verify if the match is exact or if there is any difference. To make this process fast, we calculate the Hamming Distance\cite{hamming-distance}, which is an XOR operation between two sequences. When there is no difference or only a few base modifications, the Hamming Distance will be small, and we can consider that a match is found.

\begin{figure}[htb]
    \centering
    \includegraphics[width=1\linewidth]{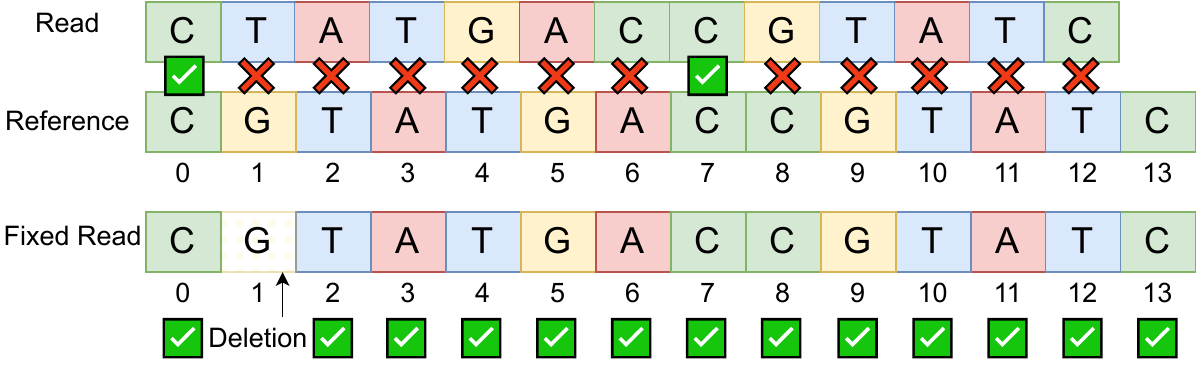}
    \caption{Illustration of how a large Hamming Distance forms when there is a deletion or insertion.}
    \label{fig:hamming-distance}
\end{figure}

However, if there is just one insertion or deletion, all the following bases can be mismatched, resulting in a huge Hamming Distance, as illustrated in \figurename~\ref{fig:hamming-distance}. Prior works\cite{genozip}\cite{seq-compress}\cite{genome-compress} consider such cases unmatchable sequences. We improve the algorithm's matching capability by further conducting a local search to calculate the Edit Distance\cite{edit-distance} when the same reference start is found but the Hamming Distance is huge. We use the WFA-2 algorithm\cite{gap-affine-alignment}\cite{wfa-2} to get the Edit Distance and the alignment CIGAR\cite{sam-format} to reconstruct the original read with insertions or deletions.

\begin{figure}[htb]
    \centering
    \includegraphics[width=0.7\linewidth]{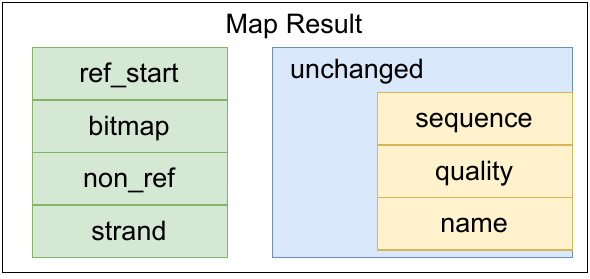}
    \caption{Map Result Fields: the quality scores, and name identifier are stored unchanged. If a match is found, the sequence can be freed. The field is for unmatched sequence and decompression. The ref\_start is an integer that indicates the matched starting position on the reference. The strand is a boolean value that indicates whether the match is a forward match or a reverse complement match. The bitmap and non\_ref together store the matching information that is necessary for reconstructing the read.}
    \label{fig:map-result}
\end{figure}

\begin{figure}[htb]
    \centering
    \includegraphics[width=1\linewidth]{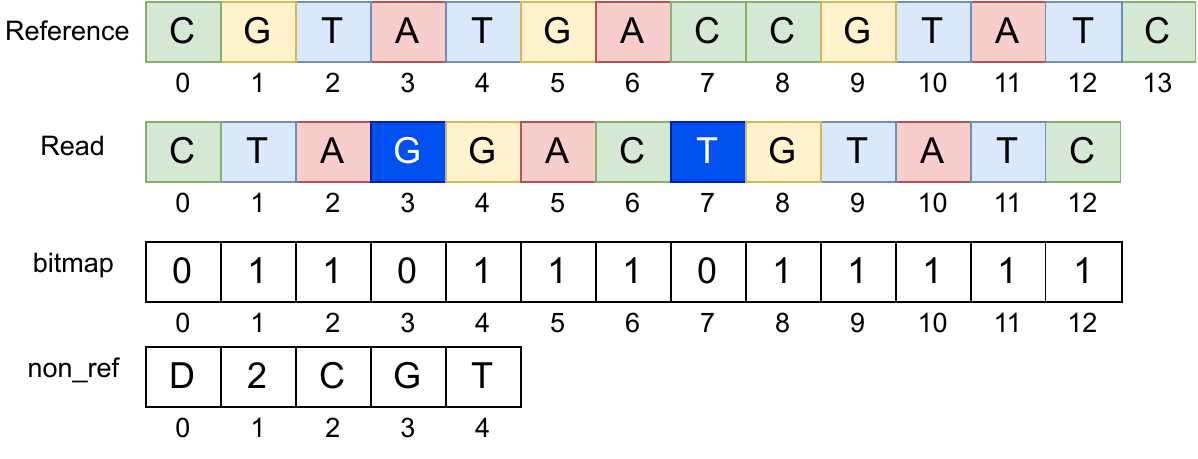}
    \caption{Bitmap and non\_ref illustration: the `G' in index 1 of reference does not exist in the read, but after that, most of the bases can be matched, so we say it is a deletion. The non\_ref will store `D' and 2 since one base is deleted. The `C' in the non\_ref is the `C' in index 0 of the read. Although it is matched to the reference, there is an indel after it, we mark it as unmatched. The two deep blue bases `G' and `T' are modifications compared to the reference, so they are marked 0 in the bitmap and recorded in the non\_ref.}
    \label{fig:bitmap-non-ref}
\end{figure}

After the sequence alignment step, each read will result in a \textit{MapResult}, the fields of which are shown in \figurename~\ref{fig:map-result}. The alignment gives us a bitmap, a non\_ref sequence, and a strand. The bitmap marks the matched and unmatched bases: for a matched base, the value is 1, and for an unmatched modification, the value is 0; if there is an insertion or deletion (indel), the position prior to the indel is marked as 0, and we store indel information in the non\_ref sequence. The non\_ref sequence stores the information for unmatched bases. If the unmatched base is a modification, non\_ref just stores the original base in the read. If the unmatched base is an indication of indel, we use `I' to mark an insertion and `D' to mark a deletion. Following an `I' or `D' is an integer number indicating how many bases are deleted or inserted plus 1, and then the base character before the indel happens, as shown in \figurename~\ref{fig:bitmap-non-ref}. With the reference sequence, ref\_start, bitmap, non\_ref, and the strand, we can fully reconstruct the original read. The data size is significantly reduced because (1) the bitmap uses a bit instead of a byte to mark a base and (2) the bitmap can be further compressed since there are usually many continuous 1s.

The alignment process reduces the sequence's size significantly but does not handle the quality scores. The quality scores have the exact same length as the sequence and thus take around half of the storage space. Because the quality scores are more random and there is no reference for such scores, we apply lossless compression methods in the segmentation step to reduce their sizes. The read names are usually short and do not take much space. We will just save them unchanged or ignore them depending on user requirements.

\subsection{Segmentation}

The segmentation process connects short map results to form a single aggregated segment for better lossless compression. The sequence and quality will be handled separately. 

In sequence segmentation, we can further reduce the reference position storage by using the difference between positions (delta) when possible. For example, when two reads have quite close reference positions---the first read's reference position is 14340909 and the second read's position is 14340997---the delta is just 88, and therefore we can use fewer bits to store the delta compared to the primary position. Moreover, for paired FASTQ files, each Map Result stores two related reads $r1$ and $r2$, which usually form a reverse complement pair. We will switch the forward read's result to $r1$ so there is a higher chance for delta to be valid in the following segmentation process.

\begin{figure}[htb]
    \centering
    \includegraphics[width=1\linewidth]{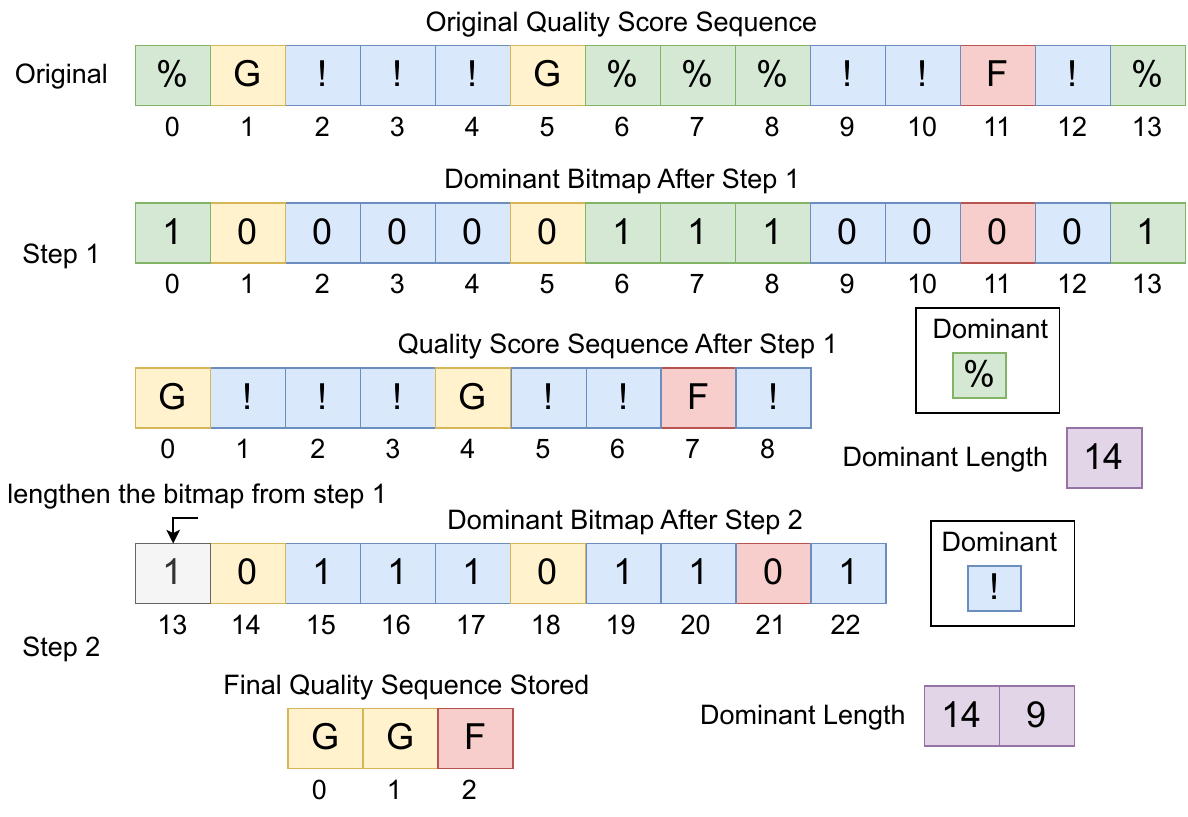}
    \caption{Dominant quality bitmap generation: when a quality score is dominant over others, we use 1 to mark them and remove them in the quality score sequence. We continue to find a dominant quality score in the remaining sequence and repeat the process. In the end, only a few non-dominant qualities will remain in the sequence. We store a bitmap, a dominant length array, a dominant quality array, and the remaining quality sequence.}
    \label{fig:dominant-quality}
\end{figure}

For quality segmentation, we use bin-quantization, dominant bitmaps, or Huffman coding to reduce their size. In each Map Result, the quality scores are a string of the read's length. According to the FASTQ format, there are 94 possible characters (from 0x21 to 0x7e) in total for each score. We propose a dominant bitmap solution as illustrated in \figurename~\ref{fig:dominant-quality} to handle this situation. The idea is to use a bit instead of a byte to store each dominant quality and let the dominant quality be further compressed by a lossless compression algorithm such as the run-length algorithm. Moreover, it is possible to cluster the scores together to form fewer quality scores if the user allows a less fine-grained quality. We call this method bin-quantization. In reality, sequencing platforms such as Illumina\cite{illumina} only provide fewer than 10 different quality scores. In this case,  Huffman coding has excellent compression performance and is fast to complete.

\subsection{Lossless Compression}

Many fields in our segmentation process can be further compressed by general-purpose lossless compression algorithms such as Zstd\cite{zstd} and Zpaq\cite{zpaq}. The lossless compression mainly deals with repeated patterns such as a long sequence of 1s or 0s in our bitmaps. Since these compressors compress a stream of bytes, we consider them as a black box to reduce field sizes. However, it is worth noting that these compressors have to store some additional header information during compression and thus do not necessarily reduce the sizes for certain fields. Therefore, we need to be wise in selecting compressors or ignoring any compressors when dealing with different fields.

\begin{figure}[htb]
    \centering
    \includegraphics[width=0.9\linewidth]{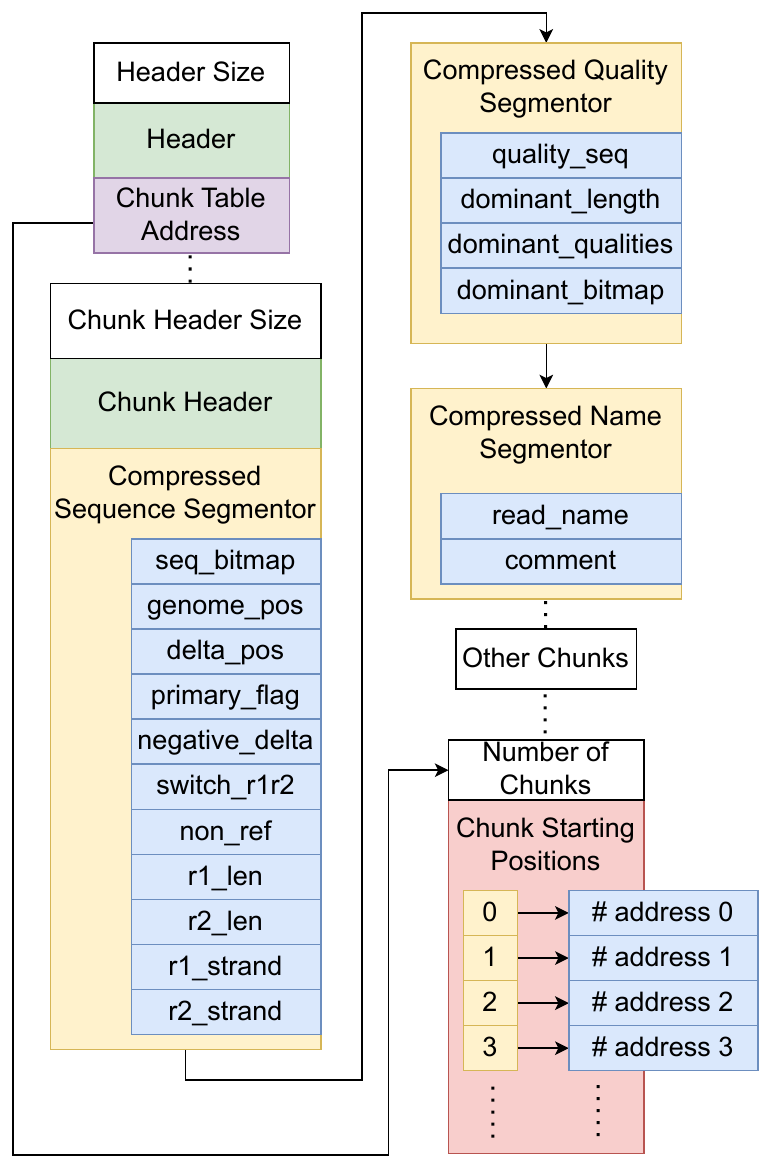}
    \caption{Compressed file structure: each chunk independently compresses its content, and provide a chunk total size to the main thread. The main thread will record each chunk's starting position, and save a table at the end of the file.}
    \label{fig:file-storage}
\end{figure}

We design a compressor selection process on a test chunk to determine which compressor with which level best fits a certain field. The selection process takes both the compression time and compression ratio into consideration. It calculates a score based on each compressor's performance on the test chunk and selects the best compressor for each field. This selection process is pure overhead for the overall compression, and thus, we use a relatively small test chunk and fix the compressor selection for the actual compression process. During our evaluations, we found that, in general, Zstd\cite{zstd} is most suitable for the sequence segmenter's field, and BSC\cite{bsc} is most suitable for the quality segmenter's field when considering both compression time and ratio. Zpaq\cite{zpaq} is the best at compression ratio, but it is several times slower compared to other compressors.

We illustrate our final compressed file structure in \figurename~\ref{fig:file-storage}. The file header has a fixed length and will be written at the beginning of the compression. It stores metadata such as the FastqZip version, whether the read names are kept, whether Gzip\cite{gzip} will be used in decompression, the sequence mode (single or paired), and so on. The chunk table address can only be known after all chunks have been compressed and written to disk. The writer thread will write the chunk table and then move the file pointer back to write the chunk table address. After the chunk table address is written, the whole compressed file is successfully stored on disk. Each chunk has its own header so that it can be independent of other chunks for better parallelism. The chunk header stores the number of elements and the compressed size for each field. The sequence segment has 11 fields, while the quality segment has four fields or just one field if Huffman coding is used to replace the dominant bitmap solution. The name segment stores the read names and comments if the read names are selected to be kept.

%% file: sections/5-evaluations.tex
\section{Evaluation}
\label{sec:evaluation}

In this section, we present the testbed and performance evaluation results of our reference-based genome compression algorithms. We evaluate the (de)compression time and compression ratio on several real-world genome sequencing datasets and compare the performance against four state-of-the-art algorithms.

\subsection{Experimental Settings}
We evaluate methods on four datasets from publicly available genome sequencing experiments: CNP0003660\cite{CNP0003660}, CNP0003664\cite{CNP0003664}, CNX0048124\cite{CNX0048124},  CNX0547764\cite{CNX0547764}, and on a standard dataset BGISEQ500\cite{BGISEQ500} published by the National Library of Medicine. Each contains sequencing results for the human genome sample NA12878 on a different platform and/or with different sequencing lengths. All results are stored in FASTQ files, and each dataset contains a pair of FASTQ files. We use the same reference sequence for all datasets during (de)compression. A more detailed description of the datasets is in \tablename~\ref{tab:datasets-description}.

\begin{table}[htb]
\centering
\caption{Genome Datasets For Compression}
\begin{tabular}{|l|l|l|l|l|}
\hline
\textbf{Dataset} & \textbf{Platform} & \textbf{Length} & \textbf{Size}  \\ \hline
E100024251\_L01\_104\cite{CNP0003660}             & DNBSEQ-T7         & PE150           & 18+20GB                 \\ \hline
CL100076243\_L01\cite{BGISEQ500}            & BGISEQ-500        & PE100           & 54+55GB               \\ \hline
E100030471QC960\_L01\cite{CNX0547764}             & DNBSEQ-T7         & PE100           & 50+52GB               \\ \hline

MGIEasyRNA4\cite{CNX0048124}             & DNBSEQ-G400       & PE150           & 5.5+5.7GB                 \\ \hline
S200032449\_L01\cite{CNP0003664}            & DNBSEQ-G50       & PE100           & 28+27GB               \\ \hline
\end{tabular}
\label{tab:datasets-description}
\end{table}

For time evaluations, we conduct all (de)compression operations on the Elastic Computing Services (ECS) on Alibaba Cloud as shown in  \tablename~\ref{tab:ecs-details}. As our algorithm exploits multi-threading, we evaluate it on ECS instances with different number of cores and analyze its scalability.

\begin{table}[htb]
\centering
\caption{ECS Machine Descriptions}
\begin{tabular}{|l|p{2.4cm}|l|p{2cm}|}
\hline
\textbf{ECS Type} & \textbf{CPU}                                             & \textbf{Memory} & \textbf{Disk}   \\ \hline
ecs.c7se.4xlarge   & 16 cores, Intel(R) Xeon(R) Platinum 8369B CPU @ 2.90GHz & 32GB           & NAS Maximum 100MB/s \\ \hline
ecs.g7.32xlarge    & 128 cores, Intel(R) Xeon(R) Platinum 8369B CPU @ 2.70GHz & 512GB           & NAS Maximum 100MB/s, ESSD PL3 Maximum 4000MB/s \\ \hline
\end{tabular}
\label{tab:ecs-details}
\end{table}

\textbf{ecs.c7se.4xlarge} is one of the most cost-effective machines available on Alibaba Cloud, and thus, we perform most of our evaluations on this machine. The NAS disk is much cheaper compared to the ESSD. However, the disk speed can be a bottleneck when using more CPU cores. Therefore, we tested on a more powerful machine with 128 CPU cores and 512GB memory to find the scalability limit. We report the results of more detailed evaluations later in this section.

\subsection{Compression Performance Evaluation}

For compression performance evaluation, we record the compression ratio (CR), (de)compression CPU time, and (de)compression wall time on five datasets, and compare FastqZip with the state-of-the-art genome sequence compression algorithms.

\begin{figure}[htb]
    \centering
    \includegraphics[width=1\linewidth]{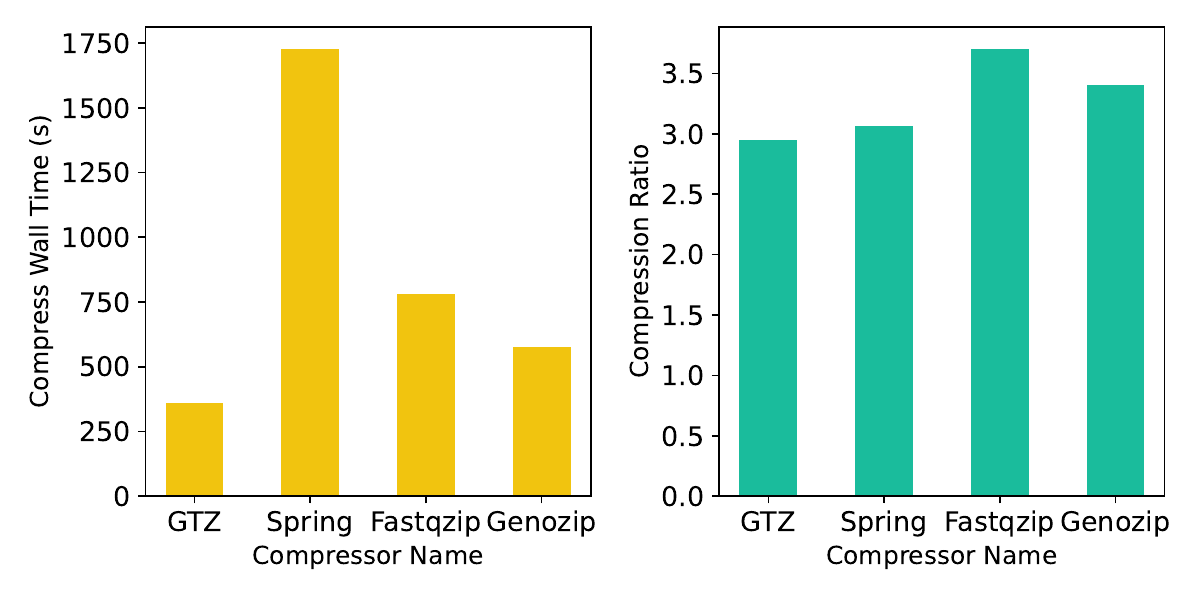}
    \caption{Compression time and ratio comparison for FastqZip and three other compression algorithms. The dataset is the first file of E100024251\_L01\_104, the compression is lossless, and each algorithm is specified to use 16 threads on the \textbf{ecs.c7se.4xlarge} machine.}
    \label{fig:comparison-compression}
\end{figure}

We run the compression on a single dataset to roughly select the compressors to compare with. As shown in \figurename~\ref{fig:comparison-compression}, Spring\cite{Spring} compresses more slowly and achieves a lower compression ratio than the other methods, and thus we do not consider it further in the following. GTZ\cite{gtz} is fast but has a lower compression ratio. It has a bug in paired FASTQ file compression and we could not get the customer support in a timely manner. We also tested FQSqueezer\cite{FQSqueezer}, but it requires more than 100GB memory and over 5000s compression time for a 300M small FASTQ file. We then tested it on the 512GB memory machine, and it stuck forever with lower than 5\% of the CPU utilization for the E100024251\_L01\_104 dataset. It is clearly too slow compared to other algorithms, and therefore, we give up on it. The LW-FQZip2\cite{LW-FQZip2} has errors in building their program. We fixed their compiling errors but it still has segmentation errors when running compression. Also, they did not provide a way to specify the output filename which is super hard to use in an automated pipeline. After much trial and error, we conclude that the best existing genome compression algorithm is Genozip\cite{genozip} in terms of project completeness, ease of use, compression ratio, and compression speed. We present detailed comparison results with Genozip later in this section.


\begin{table*}[htb]
\centering
\caption{Lossless compression performance comparison against Genozip}
\begin{tabular}{|l|llll|}
\hline
\multicolumn{1}{|c|}{\textbf{Tool}}                                                   & \textbf{Dataset}     & \textbf{CR}    & \textbf{Compress CPU Time} & \textbf{Decompress CPU Time} \\ \hline
\multirow{5}{*}{\begin{tabular}[c]{@{}l@{}}FastqZip\\ Quality\\ Huffman\end{tabular}} & E100024251\_L01\_104 & 2.93           & \textbf{151m48s}       & \textbf{94m20s}          \\
                                                                                      & MGIEasyRNA4          & 1.91           & 175m6s                 & \textbf{26m19s}          \\
                                                                                      & S200032449\_L01      & 2.32           & 163m8s                 & 148m9s                   \\
                                                                                      & CL100076243\_L01     & 2.23           & \textbf{417m17s}       & \textbf{251m15s}         \\
                                                                                      & E100030471QC960\_L01 & 2.30           & \textbf{522m16s}       & \textbf{245m31s}         \\ \hline
\multirow{5}{*}{\begin{tabular}[c]{@{}l@{}}FastqZip\\ Dominant\\ Bitmap\end{tabular}} & E100024251\_L01\_104 & \textbf{3.37}  & 237m31s                & 167m36s                  \\
                                                                                      & MGIEasyRNA4          & \textbf{2.04}  & 213m50s                & 55m42s                   \\
                                                                                      & S200032449\_L01      & \textbf{2.33}  & 378m32s                & 295m27s                  \\
                                                                                      & CL100076243\_L01     & \textbf{2.44}  & 827m17s                & 517m9s                   \\
                                                                                      & E100030471QC960\_L01 & \textbf{2.54}  & 905m55s                & 489m40s                  \\ \hline
\multirow{5}{*}{Genozip}                                                              & E100024251\_L01\_104 & 3.14           & 160m28s                & 100m14s                  \\
                                                                                      & MGIEasyRNA4          & 2.02           & \textbf{88m46s}        & 33m5s                    \\
                                                                                      & S200032449\_L01      & 2.30           & \textbf{151m1s}        & \textbf{125m7s}          \\
                                                                                      & CL100076243\_L01     & 2.33           & 572m5s                 & 303m54s                  \\
                                                                                      & E100030471QC960\_L01 & 2.45           & 526m43s                & 281m14s                  \\ \hline
\end{tabular}
\label{tab:compression-comparison}
\end{table*}

As shown in \tablename~\ref{tab:compression-comparison}, our algorithm has a better compression ratio in all datasets compared to Genozip while having a slightly slower (de)compression time. The CR is the compression ratio calculated against the gzipped FASTQ files. We show that the compression is slower mainly due to the quality score compression--- by using Huffman coding on the quality scores, our algorithm has a faster (de)compression time over most of the datasets. The reason lies in the length of the dominant bitmap. It can be extremely long when there are multiple rounds of dominant quality selections. We need to iterate through longer than the whole quality score sequence to mark `1's and `0's in the bitmap. More importantly, the lossless compression algorithm needs to go over such a long sequence to find patterns and further compress them. This process can be either faster with a lower compression ratio or slower with a higher compression ratio. We applied the BSC\cite{bsc} algorithm for the final lossless compression to achieve a high compression ratio with a relatively short amount of time. We also tried applying the same BSC algorithm to the Huffman coding. The time cost and compressed size would be very similar for these two FastqZip approaches: the Huffman method would be around 1.5\% faster, while the compression ratio has only a difference under 0.5\% between the two. If we change this algorithm to ZPAQ\cite{zpaq}, the compression ratio can get slightly higher but it requires significantly longer compression time.

\begin{table*}[htb]
\centering
\caption{Time cost in seconds for each stage in the FastqZip algorithm}
\begin{tabular}{|l|lrrrr|}
\hline
\textbf{Tool} & \textbf{Dataset}     & \textbf{Index} & \textbf{Align} & \textbf{Sequence} & \textbf{Quality} \\ \hline
\multirow{5}{*}{\begin{tabular}[c]{@{}l@{}}FastqZip\\ Dominant\\ Bitmap\end{tabular}}      & E100024251\_L01\_104 & 7                & 7013                 & 417                    & 5276                  \\
      & MGIEasyRNA4          & 10               & 9784                 & 315                    & \num{2642}                  \\
        & S200032449\_L01      & 12               & 8487                 & 614                    & \num{10687}                 \\
              & CL100076243\_L01     & 11               & 23                & 1363                   & 20                 \\
              & E100030471QC960\_L01 & 12               & 28                & 1411                   & 24                 \\ \hline
\multirow{5}{*}{\begin{tabular}[c]{@{}l@{}}FastqZip\\ Quality\\ Huffman\end{tabular}}      & E100024251\_L01\_104 & 7                & 5998                 & 350                    & 1413                  \\
       & MGIEasyRNA4          & 12               & 9725                 & 300                    & 370                   \\
       & S200032449\_L01      & 12               & 6338                 & 484                    & 814                   \\
              & CL100076243\_L01     & 14               & 18                & 1071                   & 1975                  \\
              & E100030471QC960\_L01 & 15               & 22                & 1060                   & 4696                  \\ \hline
\end{tabular}
\label{tab:time-cost-each-stage}
\end{table*}

\begin{table*}[htb]
\centering
\caption{Original size and compression \% for sequence and quality data when losslessly compressed with XXX}
\begin{tabular}{|l|lrrrr|}
\hline
& & \multicolumn{2}{c}{\textbf{Sequence data}} & \multicolumn{2}{c|}{\textbf{Quality scores}} \\
\textbf{Tool} & \textbf{Dataset}     & \textbf{Size (GB)} & \textbf{Compression \%} & \textbf{Size (GB)} & \textbf{Compression \%} \\ \hline
\multirow{5}{*}{\begin{tabular}[c]{@{}l@{}}FastqZip\\ Dominant\\ Bitmap\end{tabular}} & CL100076243\_L01     & 5.63         & 11.87                   & 41.8                & 88.13             \\
                                                                                      & E100024251\_L01\_104 & 1.32         & 11.29               & 10.3                     & 88.71             \\
                                                                                      & E100030471QC960\_L01 & 4.54       & 10.61                      & 38.2                  & 89.39             \\
                                                                                      & MGIEasyRNA4          & 1.02           & 17.41             & 4.82                    & 82.59             \\
                                                                                      & S200032449\_L01      & 2.37           & 9.62            & 22.3                      & 90.38             \\ \hline
\multirow{5}{*}{\begin{tabular}[c]{@{}l@{}}FastqZip\\ Quality\\ Huffman\end{tabular}} & CL100076243\_L01     & 5.63           & 10.84               & 46.3                   & 89.16             \\
                                                                                      & E100024251\_L01\_104 & 1.32            & 9.82               & 12.1                 & 90.18             \\
                                                                                      & E100030471QC960\_L01 & 4.54          & 9.63                 & 42.6                  & 90.37             \\
                                                                                      & MGIEasyRNA4          & 1.02              & 16.30             & 5.22                  & 83.70             \\
                                                                                      & S200032449\_L01      & 2.37              & 9.57                & 22.4                & 90.43             \\ \hline
\end{tabular}
\label{tab:sequence-quality-size-portion}
\end{table*}

\begin{figure}[htb]
    \centering
    \includegraphics[width=1\linewidth]{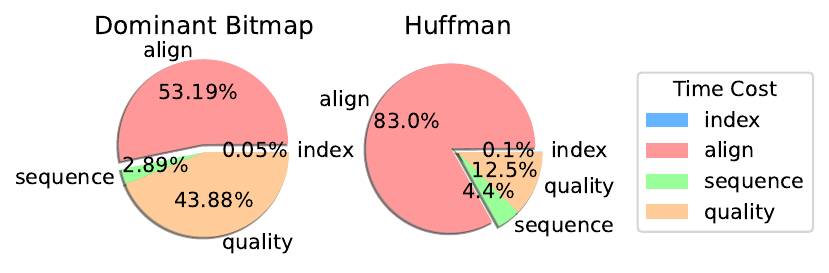}
    \caption{The time cost percentage for each stage in the \textbf{lossless} FastqZip compression pipeline: `index' represents the index loading stage, `align' represents the sequence alignment stage, `sequence' represents the sequence segmentation stage, and `quality' represents the quality segmentation stage.}
    \label{fig:stages-pct-pie}
\end{figure}

\begin{table*}[htb]
\centering
\caption{Size reduction achieved via quality score compression after applying bin clustering}
\begin{tabular}{|l|lcccc|}
\hline
 &  & \textbf{Original} & \textbf{Compressed} & \textbf{Reduction} & \textbf{Overall} \\
\textbf{Tool}  & \textbf{Dataset} & \textbf{Size} & \textbf{Size} & \textbf{ \%} & \textbf{CR} \\ \hline
\multirow{5}{*}{\begin{tabular}[c]{@{}l@{}}FastqZip\\ Dominant\\ Bitmap\end{tabular}}         & CL100076243\_L01     & 4.18E+10          & \textbf{1.70E+10}      & 59.43             & 5.13                \\
                                                                                              & E100024251\_L01\_104 & 1.03E+10          & \textbf{3.04E+09}      & 70.61             & 9.01                \\
                                                                                              & E100030471QC960\_L01 & 3.82E+10          & \textbf{1.30E+10}      & 66.11             & 6.20                \\
                                                                                              & MGIEasyRNA4          & 4.82E+09          & \textbf{1.90E+09}      & 60.67             & 4.09                \\
                                                                                              & S200032449\_L01      & 2.23E+10          & \textbf{1.01E+10}      & 54.45             & 4.60                \\ \hline
\multirow{5}{*}{\begin{tabular}[c]{@{}l@{}}FastqZip\\ Quality\\ Huffman\\ + BSC\end{tabular}} & CL100076243\_L01     & 4.63E+10          & \textbf{1.72E+10}      & 62.75             & 5.06                \\
                                                                                              & E100024251\_L01\_104 & 1.21E+10          & \textbf{3.08E+09}      & 74.56             & 8.94                \\
                                                                                              & E100030471QC960\_L01 & 4.26E+10          & \textbf{1.30E+10}      & 69.45             & 6.18                \\
                                                                                              & MGIEasyRNA4          & 5.22E+09          & \textbf{1.92E+09}      & 63.31             & 4.10                \\
                                                                                              & S200032449\_L01      & 2.24E+10          & \textbf{1.01E+10}      & 54.75             & 4.60                \\ \hline
\end{tabular}
\label{tab:cluster-quality-bin}
\end{table*}

We also evaluate the time cost and compressed size for each stage to find out whether the computing resources are spent on the right part. The results are recorded in \tablename~\ref{tab:time-cost-each-stage} and \tablename~\ref{tab:sequence-quality-size-portion}. As shown in \figurename~\ref{fig:stages-pct-pie}, the aligner and quality segmentation stages take the majority amount of time during compression for our dominant-quality algorithm, while the Huffman algorithm introduces very little time for quality score compression and thus makes the alignment takes over 80\% of the time. Moreover, \tablename~\ref{tab:sequence-quality-size-portion} shows that the quality scores actually take up the majority of space (over 80\%) in the compressed file. In theory, because of our index structure and local alignment mechanism, our alignment process would take longer compared to Genozip's method. A longer search time and more thorough sequence comparison against the reference lead to a higher compression ratio for the sequence part. The result indicates that we have done a great job in sequence compression. To further reduce the file sizes, it makes more sense for us to focus on reducing the size of the quality scores. But because the quality scores consisted of relatively random values, there is no reference for such values to be compactly compressed. Since most downstream analysis programs only need the sequence values or do not require super accurate quality scores, we refer to lossy compression for a higher compression ratio.

\begin{figure}[htb]
    \centering
    \includegraphics[width=1\linewidth]{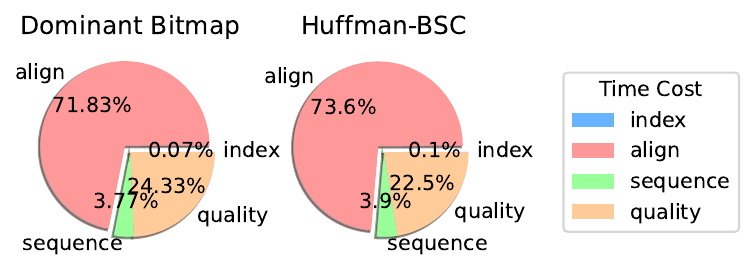}
    \caption{The time cost percentage for each stage with \textbf{lossy} quality scores (clustered to 8 bins). The Huffman-BSC means a BSC algorithm is applied after Huffman.}
    \label{fig:lossy-quality-pie}
\end{figure}

We conduct lossy compression to quality scores by grouping quality scores into fewer classes. ``N bins'' means we select the N most popular quality scores and transform all the rest quality scores into the N scores according to their distance. \figurename~\ref{fig:lossy-quality-pie} shows that by applying quality score clustering, the dominant bitmap method spends less time on compressing the quality scores, resulting in faster compression. \tablename~\ref{tab:cluster-quality-bin} shows that the quality bin clustering is effective in both algorithms and can reduce more than 50\% of storage space requirement. By applying a BSC compression after the Huffman coding, these two approaches have very similar compression performance. The similar performance also suggests that we probably have already reached the limit of quality score compression. Other than using very few quality bins (the extreme case is just one bin, and we basically ignore the quality scores) or obtaining some additional information from the sequencing process(like if there is some pattern for the quality score distribution using some specific sequencing method), there is not much room to improve the compression ratio over the quality scores.

\subsection{Resource Consumption Analysis}

The memory and CPU usage of FastqZip is mainly determined by the number of threads and the read number for each chunk. It is important to know the memory usage characteristics to give an appropriate setting. An unmatched setting can lead to memory allocation failure or underutilization of CPU resources. In this section, we evaluate the memory consumption and CPU utilization of FastqZip to better understand how we should set these two hyperparameters to maximize the compression performance. 

\begin{figure}[htb]
    \centering
    \includegraphics[width=1\linewidth]{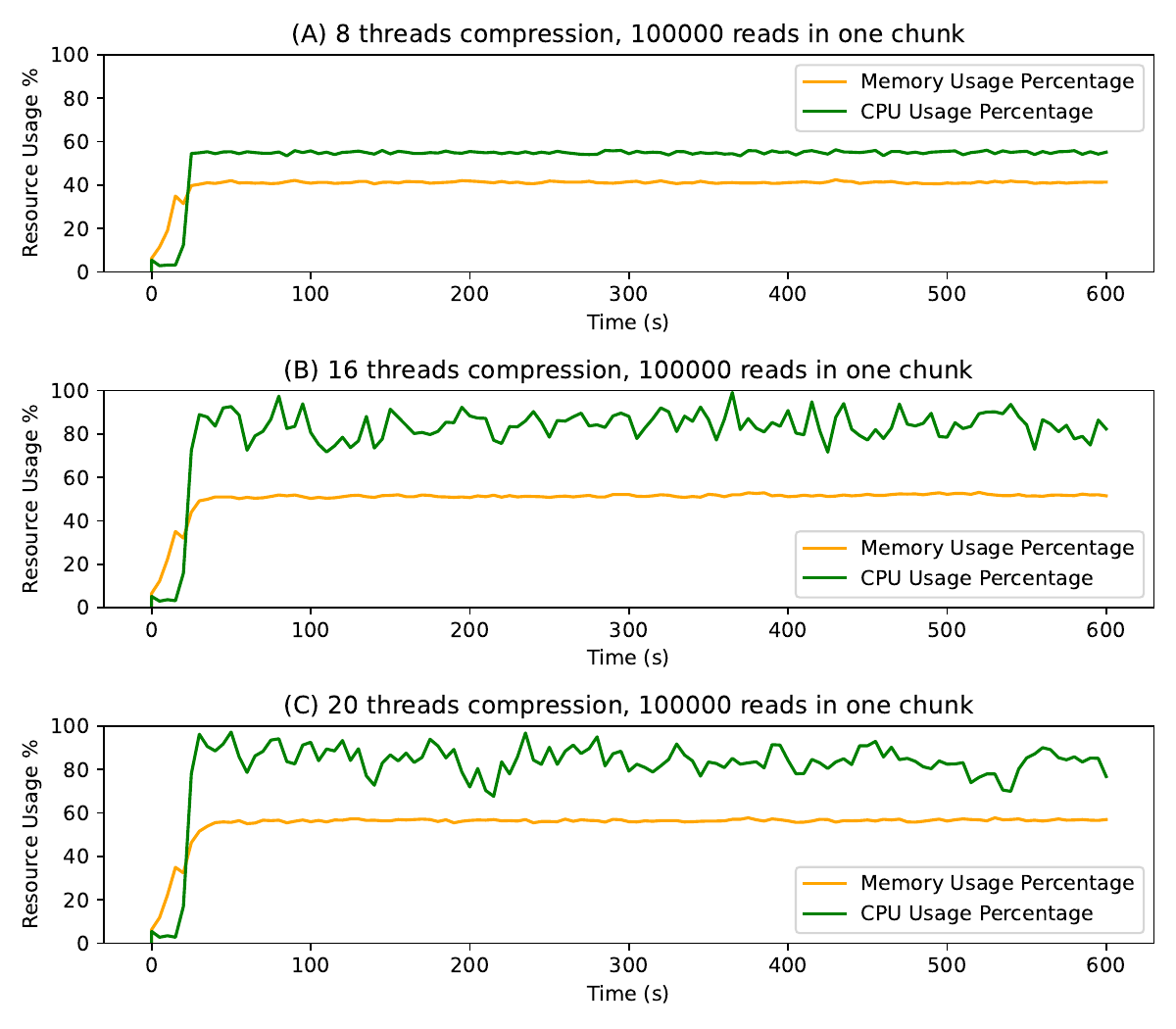}
    \caption{Memory and CPU usage during FastqZip compression. Tests are run on a \textbf{ecs.c7se.4xlarge} cloud server with 16 CPUs and 32GB memory; the dataset is E100024251\_L01\_104.}
    \label{fig:memory-cpu}
\end{figure}

Our experiments show that an appropriate read number setting can ensure that memory consumption does not exceed the physical bound. As shown in \figurename~\ref{fig:memory-cpu}, by setting the read number of each chunk to 100,000, the memory consumption is between 50\% and 60\% (around 19GB) even if we utilize all CPU resources. If we have less memory---say 16GB---we can either reduce the number of threads or set a smaller read number per chunk. A smaller read number can ensure that the program does not fail due to memory allocation errors, but can also increase the total number of chunks, resulting in a slight increase in compressed size due to the chunk header. But because the chunk header size is usually negligible compared to the chunk content size, we usually do not need to optimize the compression ratio through read numbers--- the chunk header is fixed to 100 bytes, while the chunk content consists of tens of thousands of reads. To be more specific, we assume putting 10,000 reads in one chunk. The 10,000 reads contain 1,000,000 bases and quality scores, and even with a (typically unrealistically high) compression ratio of 50, the chunk content would still be around 40,000 bytes and the chunk header would take just 100/40000 = 0.25\% of the space. When we compress 100,000 reads in a chunk, the same assumed compression ratio yields a header taking only 0.025\% of the space. Nonetheless, when the memory is enough, we recommend a larger chunk for this small optimization with no harm.

On the other hand, our algorithm is capable of fully utilizing computing resources with an appropriate number of threads. \figurename~\ref{fig:memory-cpu} (B) shows that by setting the thread number to 16 (the number of available CPUs), each worker can take up one CPU core and reach over 90\% of CPU utilization. The slight oversubscription shown in \figurename~\ref{fig:memory-cpu} (C) does not increase the CPU utilization more while having some troughs that drop to 65\% of utilization. We can safely conclude that setting the thread number to the number of CPU cores available can utilize the computing resources well enough.

\subsection{Scalability Evaluation}

When we increase the number of CPUs for (de)compression, the (de)compression wall time should decrease and the throughput should climb up because of parallelism. But at some point, the I/O can be a bottleneck. Because we need to decompress the Gzipped raw file to read the content before compression, the read thread may soon be too slow to get the data for hundreds of worker threads to do the compression parallelly. In this section, we evaluate the parallel scalability of FastqZip on the selected Alibaba Elastic Computing Services to see if the (de)compression wall time can continue decreasing with more and more CPU cores.

\begin{figure}[htb]
    \centering
    \includegraphics[width=1\linewidth]{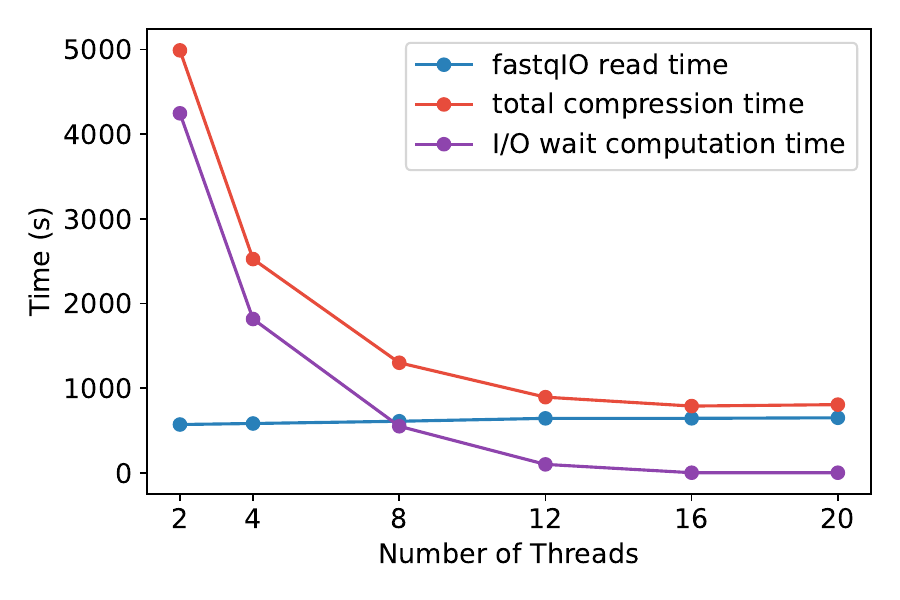}
    \caption{Scalability evaluation of FastqZip: The evaluation is performed on \textbf{ecs.g7.32xlarge} with sufficient CPUs and memory.}
    \label{fig:scaling-fastqzip}
\end{figure}

As shown in \figurename~\ref{fig:scaling-fastqzip}, FastqZip scales pretty well when there are fewer than 16 threads. The I/O wait computation time decreases to 0 when there are 20 threads, meaning the I/O has been too slow to provide enough data for so many threads to consume. If the I/O is faster than the computation, the read buffer will build up to full and the I/O has to wait for the computation to finish to continue reading data. The total compression time converges with fastqIO read time when enough threads are provided. No further speed-up is possible other than having a faster I/O because the limit is to read all the data in and the compression just finishes in no time.

%% file: sections/6-conculsions.tex
\section{Conclusions \& Future Work}
\label{sec:conclusion}

We developed a novel genome sequence compression framework FastqZip, which consists of indexing, sequence alignment, segmentation, and lossless compression modules. We proposed a better sequence alignment approach that matches the reads with insertions and deletions so that more reads can be compressed by aligning them to the reference sequence. Based on our evaluations of five real-world DNA and RNA sequencing datasets, we report the following key findings:

\begin{itemize}
    \item The proposed alignment search algorithm can further reduce the sequence size in the compressed file and reaches a compression ratio that is better than the state-of-the-art genome sequence compression algorithms, including Genozip\cite{genozip}.
    \item The quality scores take up over 80\% of storage space in the compressed file, and if they can be lossy compressed, the compression ratio can be around 2X higher.
    \item We use two different algorithms, Dominant Bitmap and Huffman, to compress the quality scores and they reach very similar performance when using the BSC lossless compression algorithm at the end, meaning that the compression of quality scores under the current approach has reached its limit.
    \item Our algorithm has great scalability and memory consumption characteristics that can ensure the memory consumption does not exceed the physical limit while utilizing all CPU resources.
\end{itemize}

For future work, we notice that the compression ratio and speed are heavily influenced by the selected lossless compression algorithm. For instance, ZPAQ\cite{zpaq} yields the best compression ratio but it is 10X slower, while the ZSTD\cite{zstd} algorithm is fast but has a relatively low compression ratio. We selected the BSC\cite{bsc} algorithm to serve as a compromise solution that is relatively fast speed and achieves a high compression ratio. We think there is still room to improve on the lossless compression algorithm, considering the special patterns in our bitmaps. Also, since our algorithm allows each chunk to be compressed independently, there is room to accelerate compression with a GPU. We plan to investigate GPU and FPGA acceleration in future work.

%% file: main.bbl
\begin{thebibliography}{10}
\providecommand{\url}[1]{#1}
\csname url@samestyle\endcsname
\providecommand{\newblock}{\relax}
\providecommand{\bibinfo}[2]{#2}
\providecommand{\BIBentrySTDinterwordspacing}{\spaceskip=0pt\relax}
\providecommand{\BIBentryALTinterwordstretchfactor}{4}
\providecommand{\BIBentryALTinterwordspacing}{\spaceskip=\fontdimen2\font plus
\BIBentryALTinterwordstretchfactor\fontdimen3\font minus
  \fontdimen4\font\relax}
\providecommand{\BIBforeignlanguage}[2]{{%
\expandafter\ifx\csname l@#1\endcsname\relax
\typeout{** WARNING: IEEEtran.bst: No hyphenation pattern has been}%
\typeout{** loaded for the language `#1'. Using the pattern for}%
\typeout{** the default language instead.}%
\else
\language=\csname l@#1\endcsname
\fi
#2}}
\providecommand{\BIBdecl}{\relax}
\BIBdecl

\bibitem{fastq-format}
\BIBentryALTinterwordspacing
P.~J.~A. Cock, C.~J. Fields, N.~Goto, M.~L. Heuer, and P.~M. Rice, ``{The
  Sanger FASTQ file format for sequences with quality scores, and the
  Solexa/Illumina FASTQ variants},'' \emph{Nucleic Acids Research}, vol.~38,
  no.~6, pp. 1767--1771, 12 2009. [Online]. Available:
  \url{https://doi.org/10.1093/nar/gkp1137}
\BIBentrySTDinterwordspacing

\bibitem{hernaez2019genomic}
M.~Hernaez, D.~Pavlichin, T.~Weissman, and I.~Ochoa, ``Genomic data
  compression,'' \emph{Annual Review of Biomedical Data Science}, vol.~2, pp.
  19--37, 2019.

\bibitem{gzip}
A.~Shah and M.~Sethi, ``The improvised gzip, a technique for real time lossless
  data compression,'' \emph{EAI Endorsed Transactions on Context-aware Systems
  and Applications}, vol.~6, p. 160599, 06 2019.

\bibitem{bzip2}
J.~Gilchrist, ``Parallel data compression with bzip2,'' 01 2004.

\bibitem{reference-compression}
B.~Chern, I.~Ochoa, A.~Manolakos, A.~No, K.~Venkat, and T.~Weissman,
  ``Reference based genome compression,'' \emph{IEEE Inf Theory Workshop, ITW},
  04 2012.

\bibitem{path-encoding-ref}
\BIBentryALTinterwordspacing
C.~Kingsford and R.~Patro, ``{Reference-based compression of short-read
  sequences using path encoding},'' \emph{Bioinformatics}, vol.~31, no.~12, pp.
  1920--1928, 02 2015. [Online]. Available:
  \url{https://doi.org/10.1093/bioinformatics/btv071}
\BIBentrySTDinterwordspacing

\bibitem{RENANO}
\BIBentryALTinterwordspacing
G.~D. y~{\'A}lvarez, G.~Seroussi, P.~Smircich, J.~Sotelo-Silveira, I.~Ochoa,
  and {\'A}.~Mart{\'\i}n, ``Renano: a reference-based compressor for nanopore
  fastq files,'' \emph{bioRxiv}, 2021. [Online]. Available:
  \url{https://www.biorxiv.org/content/early/2021/06/01/2021.03.26.437155}
\BIBentrySTDinterwordspacing

\bibitem{genozip}
\BIBentryALTinterwordspacing
D.~Lan, R.~Tobler, Y.~Souilmi, and B.~Llamas, ``{Genozip: a universal
  extensible genomic data compressor},'' \emph{Bioinformatics}, vol.~37,
  no.~16, pp. 2225--2230, 02 2021. [Online]. Available:
  \url{https://doi.org/10.1093/bioinformatics/btab102}
\BIBentrySTDinterwordspacing

\bibitem{initial-genome}
{International Human Genome Sequencing Consortium}, ``Initial sequencing and
  analysis of the human genome,'' \emph{Nature}, vol. 409, p. 860–921, 2001.

\bibitem{seq-compress}
\BIBentryALTinterwordspacing
M.~Sardaraz, M.~Tahir, A.~A. Ikram, and H.~Bajwa, ``Seqcompress: An algorithm
  for biological sequence compression,'' \emph{Genomics}, vol. 104, no.~4, pp.
  225--228, 2014. [Online]. Available:
  \url{https://www.sciencedirect.com/science/article/pii/S0888754314001499}
\BIBentrySTDinterwordspacing

\bibitem{genome-compress}
U.~Ghoshdastider and B.~Saha, ``Genomecompress: A novel algorithm for dna
  compression,'' 2007.

\bibitem{ngs-application}
\BIBentryALTinterwordspacing
A.~K. Gupta and U.~Gupta, ``Chapter 20 - next generation sequencing and its
  applications,'' in \emph{Animal Biotechnology (Second Edition)}, 2nd~ed.,
  A.~S. Verma and A.~Singh, Eds.\hskip 1em plus 0.5em minus 0.4em\relax Boston:
  Academic Press, 2020, pp. 395--421. [Online]. Available:
  \url{https://www.sciencedirect.com/science/article/pii/B9780128117101000185}
\BIBentrySTDinterwordspacing

\bibitem{human-genome-project}
L.~Hood and R.~L., ``The human genome project: big science transforms biology
  and medicine.'' \emph{Genome Med}, 2013.

\bibitem{maxam-gilbert}
\BIBentryALTinterwordspacing
W.~Gilbert and A.~Maxam, ``The nucleotide sequence of the <i>lac</i>
  operator,'' \emph{Proceedings of the National Academy of Sciences}, vol.~70,
  no.~12, pp. 3581--3584, 1973. [Online]. Available:
  \url{https://www.pnas.org/doi/abs/10.1073/pnas.70.12.3581}
\BIBentrySTDinterwordspacing

\bibitem{sanger-sequencing}
C.~A. Sanger~F, Nicklen~S, ``Dna sequencing with chain-terminating
  inhibitors,'' \emph{Proc Natl Acad Sci U S A.}, 12 1977.

\bibitem{illumina}
S.~R. et~al., ``High-throughput snp genotyping on universal bead arrays.''
  \emph{Mutat Res.}, pp. 70--82, 06 2005.

\bibitem{LW-FQZip}
Y.~Zhang, L.~Li, Y.~Yang, X.~Yang, S.~He, and Z.~Zhu, ``Light-weight
  reference-based compression of fastq data,'' \emph{BMC bioinformatics},
  vol.~16, pp. 1--8, 2015.

\bibitem{LW-FQZip2}
Z.-A. Huang, Z.~Wen, Q.~Deng, Y.~Chu, Y.~Sun, and Z.~Zhu, ``Lw-fqzip 2: a
  parallelized reference-based compression of fastq files,'' \emph{BMC
  bioinformatics}, vol.~18, pp. 1--8, 2017.

\bibitem{gtz}
Y.~Xing, G.~Li, Z.~Wang, B.~Feng, Z.~Song, and C.~Wu, ``Gtz: a fast compression
  and cloud transmission tool optimized for fastq files,'' \emph{BMC
  bioinformatics}, vol.~18, no.~16, pp. 233--242, 2017.

\bibitem{LEON}
G.~Benoit, C.~Lemaitre, D.~Lavenier, E.~Drezen, T.~Dayris, R.~Uricaru, and
  G.~Rizk, ``Reference-free compression of high throughput sequencing data with
  a probabilistic de bruijn graph,'' \emph{BMC bioinformatics}, vol.~16, no.~1,
  pp. 1--14, 2015.

\bibitem{Quip}
D.~C. Jones, W.~L. Ruzzo, X.~Peng, and M.~G. Katze, ``Compression of
  next-generation sequencing reads aided by highly efficient de novo
  assembly,'' \emph{Nucleic acids research}, vol.~40, no.~22, pp. e171--e171,
  2012.

\bibitem{DSRC2}
{\L}.~Roguski and S.~Deorowicz, ``Dsrc 2—industry-oriented compression of
  fastq files,'' \emph{Bioinformatics}, vol.~30, no.~15, pp. 2213--2215, 2014.

\bibitem{FQC}
A.~Dutta, M.~M. Haque, T.~Bose, C.~V. S.~K. Reddy, and S.~S. Mande, ``Fqc: A
  novel approach for efficient compression, archival, and dissemination of
  fastq datasets,'' \emph{Journal of bioinformatics and computational biology},
  2015.

\bibitem{fqzcomp}
J.~Bonfield and M.~Mahoney, ``Compression of fastq and sam format sequencing
  data,'' \emph{PloS one}, vol.~8, p. e59190, 03 2013.

\bibitem{LFQC}
\BIBentryALTinterwordspacing
M.~Nicolae, S.~Pathak, and S.~Rajasekaran, ``{LFQC: a lossless compression
  algorithm for FASTQ files},'' \emph{Bioinformatics}, vol.~31, no.~20, pp.
  3276--3281, 06 2015. [Online]. Available:
  \url{https://doi.org/10.1093/bioinformatics/btv384}
\BIBentrySTDinterwordspacing

\bibitem{Spring}
S.~Chandak, K.~Tatwawadi, I.~Ochoa, M.~Hernaez, and T.~Weissman, ``Spring: a
  next-generation compressor for fastq data,'' \emph{Bioinformatics}, Aug 2019.

\bibitem{FQSqueezer}
S.~Deorowicz, ``Fqsqueezer: k-mer-based compression of sequencing data,''
  \emph{Scientific Reports}, 2020.

\bibitem{hamming-distance}
A.~Bookstein, V.~Kulyukin, and T.~Raita, ``Generalized hamming distance,''
  \emph{Information Retrieval}, vol.~5, 10 2002.

\bibitem{edit-distance}
C.~Zhao, ``String correction using the damerau-levenshtein distance,''
  \emph{BMC Bioinformatics}, 06 2019.

\bibitem{gap-affine-alignment}
\BIBentryALTinterwordspacing
S.~Marco-Sola, J.~M. Eizenga, A.~Guarracino, B.~Paten, E.~Garrison, and
  M.~Moreto, ``{Optimal gap-affine alignment in O(s) space},''
  \emph{Bioinformatics}, vol.~39, no.~2, p. btad074, 02 2023. [Online].
  Available: \url{https://doi.org/10.1093/bioinformatics/btad074}
\BIBentrySTDinterwordspacing

\bibitem{wfa-2}
\BIBentryALTinterwordspacing
S.~Marco-Sola, J.~C. Moure, M.~Moreto, and A.~Espinosa, ``{Fast gap-affine
  pairwise alignment using the wavefront algorithm},'' \emph{Bioinformatics},
  vol.~37, no.~4, pp. 456--463, 09 2020. [Online]. Available:
  \url{https://doi.org/10.1093/bioinformatics/btaa777}
\BIBentrySTDinterwordspacing

\bibitem{sam-format}
\BIBentryALTinterwordspacing
H.~Li, B.~Handsaker, A.~Wysoker, T.~Fennell, J.~Ruan, N.~Homer, G.~Marth,
  G.~Abecasis, R.~Durbin, and .~G. P. D.~P. Subgroup, ``{The Sequence
  Alignment/Map format and SAMtools},'' \emph{Bioinformatics}, vol.~25, no.~16,
  pp. 2078--2079, 06 2009. [Online]. Available:
  \url{https://doi.org/10.1093/bioinformatics/btp352}
\BIBentrySTDinterwordspacing

\bibitem{zstd}
Facebook, ``Zstandard,'' \url{https://github.com/facebook/zstd/releases}.

\bibitem{zpaq}
M.~V. Mahoney, ``The zpaq compression algorithm,'' 2015,
  \url{https://api.semanticscholar.org/CorpusID:13248511}.

\bibitem{bsc}
\BIBentryALTinterwordspacing
M.~Sardaraz and M.~Tahir, ``Sca-ngs: Secure compression algorithm for next
  generation sequencing data using genetic operators and block sorting,''
  \emph{Science Progress}, vol. 104, no.~2, p. 00368504211023276, 2021, pMID:
  34143692. [Online]. Available:
  \url{https://doi.org/10.1177/00368504211023276}
\BIBentrySTDinterwordspacing

\bibitem{CNP0003660}
X.~Hongxin, ``{DNBSEQT7 WES-PE150} demo data,''
  \url{https://db.cngb.org/search/project/CNP0003660/}, 10 2022.

\bibitem{CNP0003664}
------, ``{MGISEQ-200 WES PE100} demo data,''
  \url{https://db.cngb.org/search/project/CNP0003664/}, 11 2022.

\bibitem{CNX0048124}
``{RNA-Seq of UHRR},'' \url{https://db.cngb.org/search/experiment/CNX0048124/}.

\bibitem{CNX0547764}
``{DNBSEQ-T7 WES PE150 },''
  \url{https://db.cngb.org/search/experiment/CNX0547764/}.

\bibitem{BGISEQ500}
``{BGISEQ500 PCRfree NA12878 CL100076243 L01},''
  \url{https://ftp-trace.ncbi.nlm.nih.gov/ReferenceSamples/giab/data/NA12878/BGISEQ500/}.

\end{thebibliography}
